\documentclass[a4paper,fleqn]{cas-sc}

\usepackage[authoryear]{natbib}

\graphicspath{{./Figures/}}

\begin{document}
\let\WriteBookmarks\relax
\def\floatpagepagefraction{1}
\def\textpagefraction{.001}
\shorttitle{Modification of the composition and density of Mercury from late accretion}
\shortauthors{R. Hyodo, H. Genda \& R. Brasser}

\title [mode = title]{Modification of the composition and density of Mercury from late accretion}

\author[1]{Ryuki Hyodo}[type=editor, orcid=0000-0003-4590-0988]
\cormark[1]
\cortext[cor1]{Corresponding author}
\address[1]{ISAS, JAXA, Sagamihara, Japan}

\author[2]{Hidenori Genda}[type=editor, orcid=0000-0001-6702-0872]

\author[2]{Ramon Brasser}[type=editor, orcid=]
\address[2]{Earth-Life Science Institute, Tokyo Institute of Technology, Tokyo 152-8550, Japan}

\begin{abstract}
Late accretion is a process that strongly modulated surface geomorphic and geochemical features of Mercury. Yet, the fate of the impactors and their effects on Mercury's surface through the bombardment epoch are not clear. Using Monte-Carlo and analytical approaches of cratering impacts, we investigate the physical and thermodynamical outcomes of late accretion on Mercury. Considering the uncertainties in late accretion, we develop scaling laws for the following parameters as a function of impact velocity and total mass of late accretion: (1) depth of crustal erosion, (2) the degree of resurfacing, and (3) mass accreted from impactor material. Existing dynamical models indicate that Mercury experienced an intense impact bombardment (a total mass of $\sim 8 \times 10^{18} - 8 \times 10^{20}$ kg with a typical impact velocity of $30-40$ km s$^{-1}$) after $4.5$ Ga. For this parameter range, we find that late accretion could remove 50 m to 10 km of the early (post-formation) crust of Mercury, but the change to its core-to-mantle ratio is negligible. Alternatively, the mantles of putative differentiated planetesimals in the early solar system could be more easily removed by impact erosion and their respective core fraction increased, if Mercury ultimately accreted from such objects. Although the cratering is notable for erasing the older geological surface records on Mercury, we show that $\sim 40-50 {\rm wt.}\%$ of the impactor's exogenic materials, including the volatile-bearing materials, can be heterogeneously implanted on Mercury's surface as a late veneer (at least $3\times 10^{18} -1.6 \times 10^{19}$ kg in total). About half of the accreted impactor's materials are vaporized, and the rest is completely melted upon the impact. We expect that the further interplay between our theoretical results and forthcoming surface observations of Mercury, including the BepiColombo mission, will lead us to a better understanding of Mercury's origin and evolution.
\end{abstract}


\maketitle

\section{Introduction}
Mercury is the densest planet in the Solar system; its metallic core is roughly 70 \% of its total mass and about twice the solar abundance \citep[e.g.,][]{Spohn2001,Hauck2013}. These values are much higher than those inferred for its sibling planets: Venus, Earth and Mars. NASA's MESSENGER (MErcury Surface Space ENvironment GEochemistry and Ranging) mission unveiled Mercury's surface to be volatile-bearing \citep{Nittler2011,Evans2015, Murchie2015, Weider2015, Peplowski2011, Peplowski2012, Peplowski2016}. The crustal K/Th and K/U ratios of Mercury are slightly higher than those of Earth, Venus, and Mars, and much higher than those of the Moon \citep{Peplowski2011,Peplowski2012}. The ratios of Cl/K and Na/Si, however, are near-chondritic \citep{Evans2012,Evans2015,Peplowski2014}. The abundance of the moderately volatile K, Na, and Cl relative to refractory Mg are most similar to those of EH enstatite chondrites \citep{Ebel2017}. \\

Several mechanisms have been proposed to explain the anomalous enhancement of the iron fraction of Mercury. Selective condensation of metal within the inner region of the solar nebula may increase the metal-to-silicate ratio of the building blocks of Mercury \citep[e.g.,][]{Weidenschilling1978}. However, this scenario requires a limited time window for the gas disk before the condensation of Mg-silicate takes place \citep{Ebel2006}. \cite{Cameron1985} suggested that evaporation of Mercury's mantle by intense heat flux from the early Sun may remove the mantle, which should lead to a substantial volatile depletion.  This is not consistent with the MESSENGER observations. Photophoresis that produces a radial drift of particles due to the thermal gradient within particles, which depends on their thermal conductivity, may separate silicate and metal \citep[e.g.,][]{Wurm2013,Loesche2016}. Detailed studies of the evolution of the inner portion of the protoplanetary disk and particle chemistry are required to validate this hypothesis. Giant impact may selectively remove the mantle of Mercury \citep[e.g.,][]{Benz1988, Benz2007, Asphaug2006, Asphaug2014}, whereas its likelihood to change its core mass fraction from 0.3 (chondritic) to 0.7 (current) in contemporary planet formation scenarios is small \citep{Chau2018, Clement2019a}. Besides, it is challenging to explain the observed volatile abundance on Mercury's surface because an energetic giant impact is expected to produce significant global melting and potential vaporization, leading to a preferential loss of volatile elements such as K while keeping refractory elements such as Th and U.\\

After the cessation of primary accretion, which may have included core-mantle separation, leftover planetesimals, asteroids and comets bombarded terrestrial planets; this intense epoch of bombardment is termed late accretion \citep[e.g.,][]{Bottke2012, Morbidelli2012, Morbidelli2018, Brasser2016, Brasser2020}. The giant impacts accompanying the last stage of primary accretion during the terrestrial planet formation process is a stochastic regime. In contrast, the late phase by numerous leftover (small) bodies is a cumulative process and has the strong potential to dictate the final global geomorphic and geochemical features of planet's surface at the very last moment of the planets' formation \citep{Melosh2011}, although a few stochastic large impacts are expected to have occurred after this time \citep{Brasser2016, Genda2017,Brasser2017}.\\

The exact timing, duration, and the dominant source (leftover planetesimals, asteroids, or comets) of late accretion are not well constrained and vary from model to model of dynamics and chronology. The early $^{40}$Ar/$^{39}$Ar analysis of Apollo samples returned from the Moon indicated a surge in the impact rate at around $\sim 3.7-3.9$ Ga, dubbed the "late lunar cataclysm", \citep[e.g.,][]{Papanastassiou1971a,Papanastassiou1971b,Turner1973}. Alternatively, crater counting of the Moon and Mars indicate a monotonic decline of the impact flux, dubbed the "accretion tail scenario" \citep[e.g.,][]{Hartmann1970, Neukum2001, Werner2015}. Different dynamical models of planet formation have been variously proposed to explain the observed data \citep[see more details in][]{Gomes2005,Marchi2009,Morbidelli2012,Morbidelli2018}. However, both the lunar samples and the crater records may inevitably have sampling biases \citep{Hartmann1975, Hartmann2003, Haskin1998, Haskin2003} and/or suffer from the problem of age resetting \citep{Boehnke2016}. Consequently, the ambiguity inherent within these different scenarios is not so straightforwardly resolved.\\

In this work, we build upon the outcomes of \cite{Mojzsis2018} with the aim to understand the degree of the impact-induced mass escape of Mercury's material, and the accretion and thermodynamical outcomes of the impactor's materials during late accretion to Mercury. These important physical-chemical parameters were not studied in the previous works. Considering the uncertainties in late accretion, we decided to generalize our arguments by parameterizing the total mass of impactors and impact conditions (the size of impactor, impact velocity, and impact angle) in our analytical and Monte-Carlo approaches. By doing this, our results can be applied to any specific bombardment scenarios/chronology and consequently used to make recommendations for observations to test these results. Then, using our generalized arguments combined with the results of existing dynamical models, we discuss the putative late accretion on Mercury and we arbitrarily define this as having occurred after 4.5 Ga to be consistent with the sampled terrestrial bodies (a.k.a. Earth, Moon, and Mars) and meteorites from the asteroid belt \citep{Mojzsis2019}.\\

\begin{figure*}
	\centering
	  \includegraphics[width=0.8\textwidth]{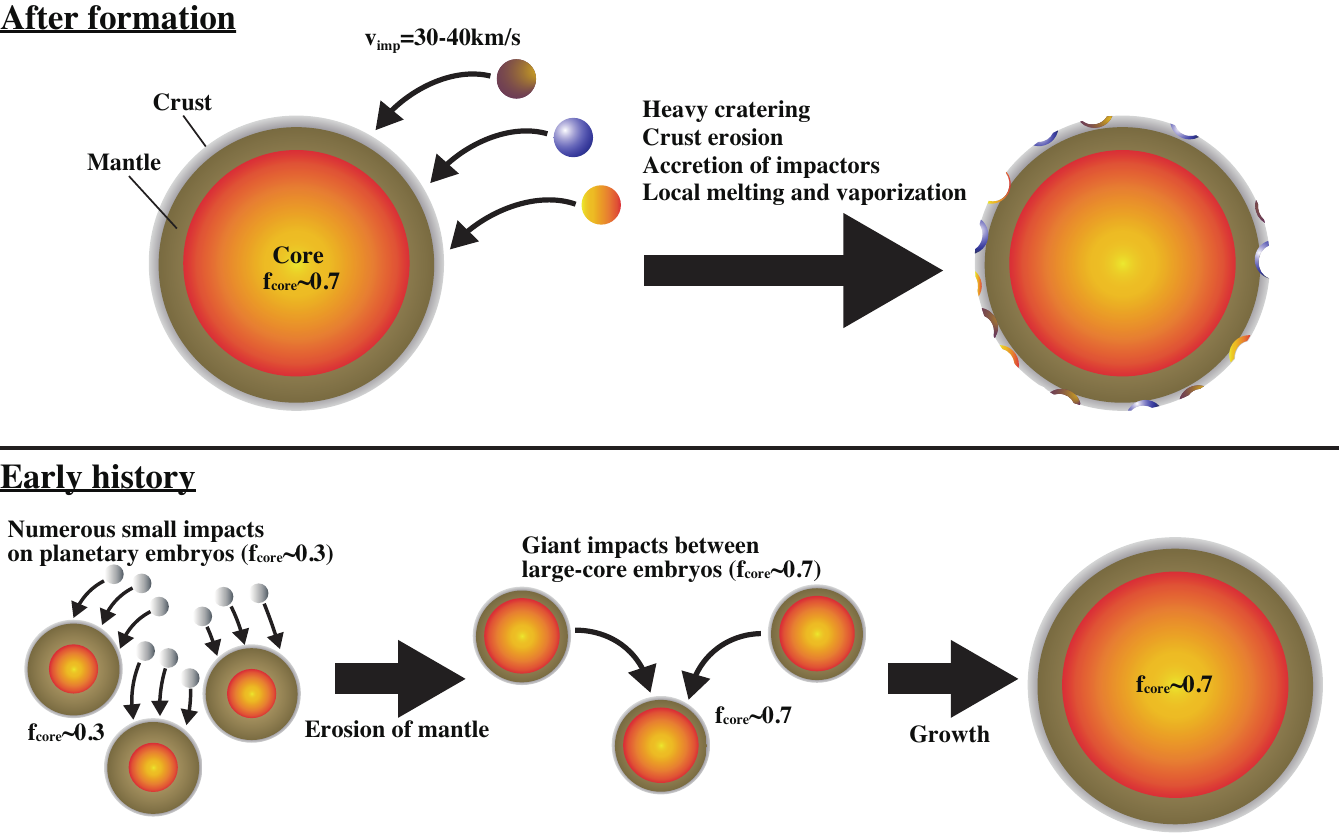}
	\caption{A summary of late accretion on Mercury after its formation studied in this work (top panel). The impactors originate from different sources (planetesimals, asteroids or comets) and have different compositions (indicated by different colors in the top panel). The typical impact velocity on Mercury is $v_{\rm imp}=30-40$ km s$^{-1}$. We show that late accretion could remove $50$ m $-10$ km of the primitive crust of Mercury depending on the total mass of the impactors of $\sim 8 \times 10^{18} - 8 \times 10^{20}$ kg (Section \ref{sec_erosion}) and produce an intensive global cratering which could erase older crustal records (Section \ref{sec_crater}). Late accretion is a process to deliver an impactor's materials to planetary bodies, and we show that about half of an impactor's materials are buried on the surface of Mercury as exogenic materials, which ought to include the volatile-bearing materials (Section \ref{sec_accretion}). About half of the accreted impactor's materials are vaporized, and the rest is completely melted upon the impact. The bombardment also induces Mercury's local crustal melting and mantle heating \citep{Mojzsis2018}.  The change in the core-to-mantle ratio during late accretion is negligible (Section \ref{sec_erosion}). Alternatively, the silicate mantles of the assumed small differentiated building blocks of Mercury could be eroded by a large fraction enlarging their core-to-mantle ratio, and Mercury might accrete from such large-core building blocks (bottom panel; Section \ref{sec_bui}).}
	\label{Fig_summary}
\end{figure*}

In section \ref{sec_model}, we summarize different models of late accretion (late bombardment) to Mercury. In section \ref{sec_method}, we describe our numerical methods for modeling late accretion on Mercury. In section \ref{sec_question1}, we show that late impact bombardment on Mercury does not contribute to a significant mass loss of the mantle of Mercury and the change in its core-to-mantle fraction is negligible, while $50$ m $-10$ km of the primordial crust of Mercury is removed depending on the total mass of impactors (top panel of Figure \ref{Fig_summary}). In section \ref{sec_question2}, we demonstrate that late accretion could produce an intensive global cratering which could erase the older geological records, while about half of the impactor's materials are buried on the surface of Mercury as exogenic materials (top panel of Figure \ref{Fig_summary}). In Section \ref{sec_bui}, we discuss an alternative scenario to produce the Mercury's large core fraction: the silicate mantles of the assumed small differentiated building blocks of Mercury could be selectively eroded hereby enlarging their core-to-mantle ratios, and it is from these objects that Mercury might accrete (bottom panel of Figure \ref{Fig_summary}). In Section \ref{sec_conclusion}, we summarize our paper.\\

\section{Models of late accretion on Mercury}
\label{sec_model}
As mentioned above, we aim to generalize our arguments in terms of impact parameters and the total mass of the impactors so that we can apply our results to arbitrary models/chronologies of late accretion and to general cratering impacts on Mercury at different times during planet formation. In this section, we briefly summarize proposed dynamical models by stating their total masses of impactors and impact velocities as reference values (see Table \ref{table_model}).\\

\subsection{Classical and sawtooth late heavy bombardment}
Here, we summarize the two cataclysm scenarios: the classical late heavy bombardment (LHB) and the sawtooth late heavy bombardment (Table \ref{table_model}). The classical late heavy bombardment includes a sudden surge of impact flux at around 3.9 Ga with its duration of $\sim100$ Myr, which was indicated by the absence of the impact signatures older than 3.9 Ga on the surface of the Moon and seemingly intense impact signatures in the lunar samples \citep[e.g.,][]{Turner1973,Tera1974}. The sawtooth-like model \citep{Morbidelli2012} was motivated to reconcile the abundance of the highly siderophile elements (HSE) in the lunar mantle \citep{Day2007,Walker2009,Walker2014,Day2015} and they argued that a sharp rise of impact flux at around 4.1 Ga with about 400 Myr decay time was likely. \cite{Mojzsis2018} calibrated these two scenarios for the mass of impactors to Mercury \citep[see also][]{Abramov2016}. They reported about $3 \times 10^{19}$ kg and $0.84 \times 10^{19}$ kg of the total impactors to Mercury with the typical impact velocity ranging from $v_{\rm imp}=30$ km s$^{-1}$ to $40$ km s$^{-1}$ for the classical and sawtooth models, respectively. Note that \cite{Morbidelli2018} updated the sawtooth model using the updated dynamical simulations and by considering the effects of the long-term magma ocean crystallization on the HSE budget in the lunar mantle \citep[see also,][]{Zhu2019}. They argued the likelihood of the accretion tail scenario and reported a factor of 10 times increase in the population of the leftover planetesimals than the previous model \citep{Morbidelli2012}, and thus the total mass of the impactor to Mercury could also increase.\\

\subsection{A recent exponential decay model}
Recently, \cite{Brasser2020} presents an updated dynamical model. In their view the late bombardment to Mercury consists of three sources: leftover planetesimals from terrestrial planet formation, the asteroid belt $-$ primarily the hypothetical E-belt or \cite{Bottke2012} $-$ and  comets arriving from the outer Solar System beyond Jupiter. Based on the abundance of highly siderophile elements (HSE) in the lunar mantle and assuming that this reservoir was the major contributor to lunar impacts shortly after its formation, \cite{Brasser2020} constrained the mass in leftover planetesimals. \cite{Day2015} use HSE abundances to argue that the Moon accreted a further 0.025 wt.\% of chondritic material after its formation while it may have still been in a purported magma ocean state. \cite{Touboul2015} and \cite{Kruijer2015} arrive at a similar amount from lunar W isotopes. Combined with the typical impact probability with the Moon the mass in leftover planetesimals was calculated to be $\sim 6 \times 10^{21}$ kg ($\sim 0.001$ Earth mass) at the time of the Moon's formation near 4.5 Ga \citep[][; cf. \cite{Connelly2016}]{Barboni2017,Thiemens2019}.\\

The ancient asteroid belt component consists mostly of the E-belt \citep{Bottke2012}. \cite{Brasser2020} showed that the E-belt and comets contributed very little to Mercury's late accretion so for simplification those two sources are ignored for this study. They reported that about $3.3^{+4.6}_{-2.6} \times 10^{20}$ kg total bombardment occurs on Mercury between 4.5 Ga and 4.0 Ga by which time the leftover population has declined by $>99$\%.\\

The impact velocity distribution is directly obtained from $N-$body simulations. The cumulative distribution of the impact velocity at Mercury was well described by using a modified Rayleigh distribution:
%
\begin{eqnarray}
	F_{\rm imp}(v_{\rm imp}) = 1 - \exp \left( \frac{-(v_{\rm imp}-\mu_{\rm imp})^{2}}{2\sigma_{\rm imp}^{2}} \right)
\label{eq_vimp}
\end{eqnarray}

\noindent where $\mu_{\rm imp} = 10.31$ km s$^{-1}$ and $\sigma_{\rm imp} = 20.81$ km s$^{-1}$, respectively. Note that we restrict the maximum velocity to 70 km s$^{-1}$ which is found by $N$-body simulations \citep[e.g.,][]{Brasser2020}. The mean value of this distribution is $<v_{\rm imp}> \sim 36 $ km s$^{-1}$. Here $\mu_{\rm imp} $ is an offset to account for a minimum impact velocity on Mercury.

\begin{table}[hbtp]
  \caption{Our analytical and Monte-Carlo approaches aim to generalize the outcome of the bombardment, such as the erosion of Mercury's surface, and the mechanical and thermodynamical fates of the impactors, for a given total mass of impactors and impact velocity. We consider the total mass of the impactor of $8 \times 10^{18} - 8 \times 10^{20}$ kg and the mean impact velocity of $v_{\rm imp} \sim 30-40$ km s$^{-1}$ as reference values to discuss the results obtained in this paper along these dynamical models.}
  \label{table_model}
  \centering
  \begin{tabular}{lcr}
  
    \hline
    Model  & Total mass of impactors to Mercury  & Impact velocity \\
    \hline \hline
  
    Decline model \\[2pt] Time: $4.5-4.0$Ga & $3.3^{+4.6}_{-2.6} \times 10^{20}$ kg \citep{Brasser2020} & Equation \ref{eq_vimp} ($<v_{\rm imp}>=36$ km $^{-1}$) \\
    \hline
  
    Classical LHB \\[2pt] Time: $3.95-3.85$Ga & $0.3 \times 10^{20}$ kg \citep{Abramov2013} & 33 km s$^{-1}$ or 43 km s$^{-1}$ \citep{Mojzsis2018}\\
    \hline
  
    Sawtooth LHB \\[2pt] Time: $4.1-3.7$Ga & $0.084 \times 10^{20}$ kg \citep{Abramov2016}  & 33 km s$^{-1}$ or 43 km s$^{-1}$ \citep{Mojzsis2018}\\
    \hline
  
  \end{tabular}
\end{table}

\section{Numerical methods for impact bombardment}
\label{sec_method}
The outcome of a single impact of a small body to a planetary body, called cratering impact, is described by the combination of parameters: impact angles, $\theta$, impactor's mass, $m_{\rm imp}$, and impact velocity, $v_{\rm imp}$ \citep[e.g.,][]{Housen2011, Melosh2011,Hyodo2020}. In this work, we use an analytical and a Monte-Carlo approach to investigate the cumulative surface erosion and impactor's accretion from late accretion after 4.5 Ga.\\

\subsection{Impact parameters}
The distribution of the impact angle is expected to have a probability density function proportional to $\sin(2\theta)$ with a peak of $\theta = 45$ degrees \citep{Shoemaker1962}, where $\theta = 90$ is the head-on collision.\\

At the size-frequency distribution (SFD) of the impactor (thus, the mass of impactors), we assume a power-law distribution as:
%
\begin{eqnarray}
	N(D) \propto D^{-\alpha}	
\label{eq_N}
\end{eqnarray}
where $D$ is the diameter of the impactor and $N(D)dD$ gives the number of bodies in a size bin of width $dD$. $\alpha$ is the slope. We use $\alpha=2$ for $D<100$ km and $\alpha=3$ for $D>100$ km \citep{Bottke2005, Masiero2015}, which are approximately the measured slopes for the present main belt SFD. The minimum size of the impactor is assumed $D_{\rm min} = 5$ km by considering a reasonable computation time. Note that our arbitrary choice of $D_{\rm min}$ does not significantly affect our results because the larger impactor carries the larger mass for $\alpha < 4$ ($M_{\rm tot,imp} \propto \int m_{\rm imp} N(D) dD \propto D^{4-\alpha}$). For $\alpha = 3$, there is an equal mass in equal size intervals (i.e., $dM_{\rm tot,imp}/dD \propto D^{3-\alpha} = {\rm constant}$ for $\alpha = 3$). The largest impactor is set to $D_{\rm max} = 1000$ km. Increasing this much further would have resulted in our impacts being dominated by a few large-mass events, which is not consistent with the cratering record.\\

The impact velocity with Mercury is a critical parameter in this work, and we use a statistical distribution (the mean of $<v_{\rm imp}>=36$ km s$^{-1}$) obtained from the latest $N$-body simulations (\cite{Brasser2020}; Equation \ref{eq_vimp}) as a reference value. In the previous papers, the impact velocity to Mercury was given $v_{\rm imp}=33$ km s$^{-1}$ \citep{Mojzsis2018} or $v_{\rm imp}=43$ km s$^{-1}$ \citep{LeDeuvre2008} obtained from a model distribution of planet crossing asteroids and comets. To study the dependence on the impact velocity, we also apply the normal distribution with the mean value of $\mu=30$ or $40$ km s$^{-1}$, and the deviation of $\sigma=1$ km s$^{-1}$.\\

\subsection{Escape of target material and accretion of impactor material}
The degree of mass escape of the target material, defined by the impact ejecta that escapes from the gravity of the target (we define this as the escape mass), and accretion of the impactor materials by cratering impacts strongly depends on impact parameters, especially the impact velocity and the mass of the target \citep[e.g.,][]{Housen2011, Melosh2011, Hyodo2020}. Based on a large number of impact simulations, \cite{Hyodo2020} developed scaling laws for the escape mass of the target material, $M_{\rm esc,tar}$, and the accretion mass of the impactor material, $M_{\rm acc,imp}$ for a wide range of impact conditions. The numerical results used to derive the scaling laws have been used and their numerical accuracies were checked in many contexts of impact phenomenon during the planet formation \citep[e.g.,][]{Kurosawa2019, Hyodo2019}. In this work, we applied these scaling laws in our Monte-Carlo simulations to track the change in mass through the mass escape of Mercury's material and the accretion of impactor's material to Mercury (Section \ref{sec_scaling}).\\

\subsection{Basic settings}
In our Monte-Carlo simulations, the impactor is represented by undifferentiated bodies with a bulk density of 3000 kg m$^{-3}$. Mercury is represented by a differentiated planet with densities of 3000 kg m$^{-3}$ and 7000 kg m$^{-3}$ for its mantle and core, respectively. The initial mass of our Mercury $M_{\rm Mer,ini}$ is a parameter, but we fix the mass of the core to that of today's ($\sim 70$wt.\% mass of current Mercury).\\

We aim to address the following two questions for late accretion on Mercury within the parameter range of the existing dynamical models: ''Does it significantly erode Mercury's mantle? (Section \ref{sec_question1})'' and ''What is the fate of impactors striking Mercury? (Section \ref{sec_question2})''. We performed $10^{5}$ times Monte-Carlo simulations for different cases of impact velocity distributions and analyze the statistical outcomes of these simulations.

\section{Does late accretion have the ability to significantly erode Mercury's mantle?}
\label{sec_question1}
In this section, we aim to understand whether late accretion could be responsible for a significant mass escape of Mercury's mantle to explain its high core-to-mantle ratio assuming the case of a smaller primordial core fraction. Below, we develop a statistical argument because the number of small impacts is numerous and outcomes of the target surface are the cumulative effects of these impacts.\\

\subsection{Escape and accretion masses weighted by the $\theta$ and $v_{\rm imp}$ distributions}
\label{sec_scaling}
The impact velocity distribution for a specific planetary body strongly depends on the age of the system and radial locations during the planet formation, while $\theta$ distribution is statistically $\sin(2\theta)$ \citep{Shoemaker1962}. \cite{Hyodo2020} derived the $\theta$-averaged escape mass of the target material, $\left< M_{\rm esc,tar}(>v_{\rm esc}) \right>_{\theta}$ and accretion mass of the impactor material, $\left< M_{\rm acc,imp}(<v_{\rm esc}) \right>_{\theta}$ in a unit of the impactor's mass as:
%
\begin{eqnarray}
	\left< \frac{M_{\rm esc,tar}(>v_{\rm esc})}{m_{\rm imp}} \right>_{\theta} =
	 0.02 \times \left( \frac{v_{\rm imp}}{v_{\rm esc}} \right)^{2.2}
	 {\rm for} \hspace{0.5em} v_{\rm imp} < 12v_{\rm esc}
\label{eq_ero_1}
\end{eqnarray}

\begin{eqnarray}
	 \left< \frac{M_{\rm esc,tar}(>v_{\rm esc})}{m_{\rm imp}} \right>_{\theta} =
	 0.076 \times \left( \frac{v_{\rm imp}}{v_{\rm esc}} \right)^{1.65}
	 {\rm for} \hspace{0.5em} v_{\rm imp} >12v_{\rm esc}
\label{eq_ero_2}
\end{eqnarray}

\begin{eqnarray}
	 \left< \frac{M_{\rm acc,imp}(<v_{\rm esc})}{m_{\rm imp}} \right>_{\theta} = 
	 0.85 - 0.071 \times \left( \frac{v_{\rm imp}}{v_{\rm esc}} \right)^{0.88}
\label{eq_acc}
\end{eqnarray}

\noindent where $v_{\rm esc}$ is the escape velocity of planet and $v_{\rm esc} \sim 4.3$ km s$^{-1}$ for Mercury. Using Equations \ref{eq_ero_1} $-$ \ref{eq_acc}, we derive the $\theta$-$v_{\rm imp}$-weighted escape and accretion masses for a given impact velocity distribution written as
%
\begin{eqnarray}
	\left< \frac{M_{\rm esc,tar}(>v_{\rm esc})}{m_{\rm imp}} \right>_{\theta,v_{\rm imp}} = \frac{ \displaystyle \int \left< \frac{M_{\rm esc,tar}(>v_{\rm esc})}{m_{\rm imp}} \right>_{\theta} \left( \frac{ dF_{\rm imp} }{dv_{\rm imp}} \right) dv_{\rm imp}}{ \displaystyle \int \left( \frac{ dF_{\rm imp} }{dv_{\rm imp}} \right) dv_{\rm imp}}
\label{eq_ero_ave}
\end{eqnarray}

\begin{eqnarray}
	\left< \frac{M_{\rm acc,imp}(<v_{\rm esc})}{m_{\rm imp}} \right>_{\theta,v_{\rm imp}} = \frac{ \displaystyle \int \left< \frac{M_{\rm acc,imp}(<v_{\rm esc})}{m_{\rm imp}} \right>_{\theta} \left( \frac{ dF_{\rm imp} }{dv_{\rm imp}} \right) dv_{\rm imp}}{ \displaystyle \int \left( \frac{ dF_{\rm imp} }{dv_{\rm imp}} \right) dv_{\rm imp}}
\label{eq_acc_ave}
\end{eqnarray}

\noindent where $F_{\rm imp}$ is the cumulative distribution of the impact velocity (for example, an offset Rayleigh in Equation \ref{eq_vimp}).\\

\begin{figure*}
	\centering
	  \includegraphics[width=\textwidth]{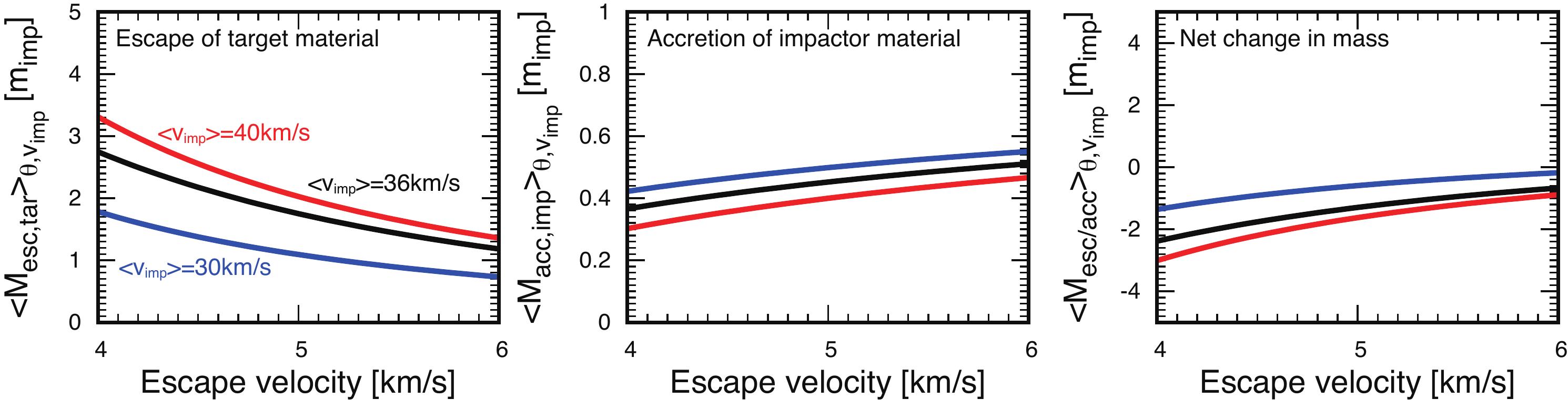}
	\caption{Escape and accretion $v_{\rm imp}$-$\theta$-weighted masses from and to planetary bodies whose escape velocities are $v_{\rm esc}=4-6$ km s$^{-1}$. Left and middle panels show the escape mass of the target material (Equation \ref{eq_ero_ave}) and the accretion mass of the impactor material (Equation \ref{eq_acc_ave}), respectively. Right panel shows the net escape/accretion mass and a negative value means a net escape and vice-versa. The blue, black and red lines represent cases where $<v_{\rm imp}>=30$, 36 and 40 km s$^{-1}$, respectively. Blue and red lines represent the cases where impact velocity distributions are a normal distribution with $\mu=30$ km s$^{-1}$ and $40$ km s$^{-1}$ with $\sigma =1$ km s$^{-1}$, respectively. The black line represents the case where a modified Rayleigh distribution (Equation \ref{eq_vimp}) is used with $\mu_{\rm imp}=10.31$ km s$^{-1}$ and $\sigma_{\rm imp} =20.81$ km s$^{-1}$ (the mean value is $<v_{\rm imp}>=36$ km s$^{-1}$).}
	\label{Fig_averaged_masses}
\end{figure*}

Figure \ref{Fig_averaged_masses} shows the $\theta$-$v_{\rm imp}$-weighted escape mass of the target material and accretion mass of the impactor material for a planet whose escape velocity is $v_{\rm esc}$ for a given impact velocity distribution ($<v_{\rm imp}>=30$, 36 and 40 km s$^{-1}$). Mass escape takes place more efficiently for a lower escape velocity at a given impact velocity distribution (left panel of Figure \ref{Fig_averaged_masses}). About half of the impactor's material accretes for $v_{\rm esc}=4-6$ km s$^{-1}$ (middle panel of Figure \ref{Fig_averaged_masses}) while the rest escapes into heliocentric orbit. The rightmost panel shows the net escape/accretion, where a negative value indicates a net escape and vice-versa (Equation \ref{eq_ero_ave} + Equation \ref{eq_acc_ave}). Between $v_{\rm esc} = 4-6$ km s$^{-1}$, net escape takes place for a mean impact velocity $v_{\rm imp} > 30$ km s$^{-1}$.\\

\subsection{Impact erosion of Mercury during late accretion}
\label{sec_erosion}
The accretion history of proto-Mercury is not well constrained \citep[][and references therein]{Ebel2017}, and high-velocity impacts generally produce mass escape of the target material. In this subsection, we consider the case where proto-Mercury has a smaller core-to-mantle ratio than that of today, assuming that the size of the core is the same as today's, that is, we assume proto-Mercury began with a larger mantle fraction. In the calculations, we do not distinguish crust and mantle by assuming they have the same density. Using the $\theta$-$v_{\rm imp}$-weighted analytical formulae (section \ref{sec_scaling}), we study whether late accretion could significantly erode Mercury's mantle and increase its core-to-mantle ratio.\\

The change in mass of Mercury's mantle during a single impact is written as:
%
\begin{eqnarray}
	\left< \frac{M_{\rm esc/acc}}{m_{\rm imp}} \right>_{\theta,v_{\rm imp}} = -\left< \frac{M_{\rm esc,tar}}{m_{\rm imp}} \right>_{\theta,v_{\rm imp}} + \left< \frac{M_{\rm acc,imp}}{m_{\rm imp}} \right>_{\theta,v_{\rm imp}} 
\label{eq_net}
\end{eqnarray}

\noindent (see Equations \ref{eq_ero_ave} and \ref{eq_acc_ave}, and Figure \ref{Fig_averaged_masses} right panel). We integrate Equation \ref{eq_net} starting from a given initial core-to-mantle fraction ($f_{\rm core,ini} < 0.7$) for a given impact velocity distribution until $f_{\rm core}$ reaches the current core mass fraction of $f_{\rm core}=0.7$. By doing this, we can estimate a required total mass of impactors to produce the current Mercury's core by erosive cratering impacts.\\

\begin{figure*}
	\centering
	  \includegraphics[width=0.6\textwidth]{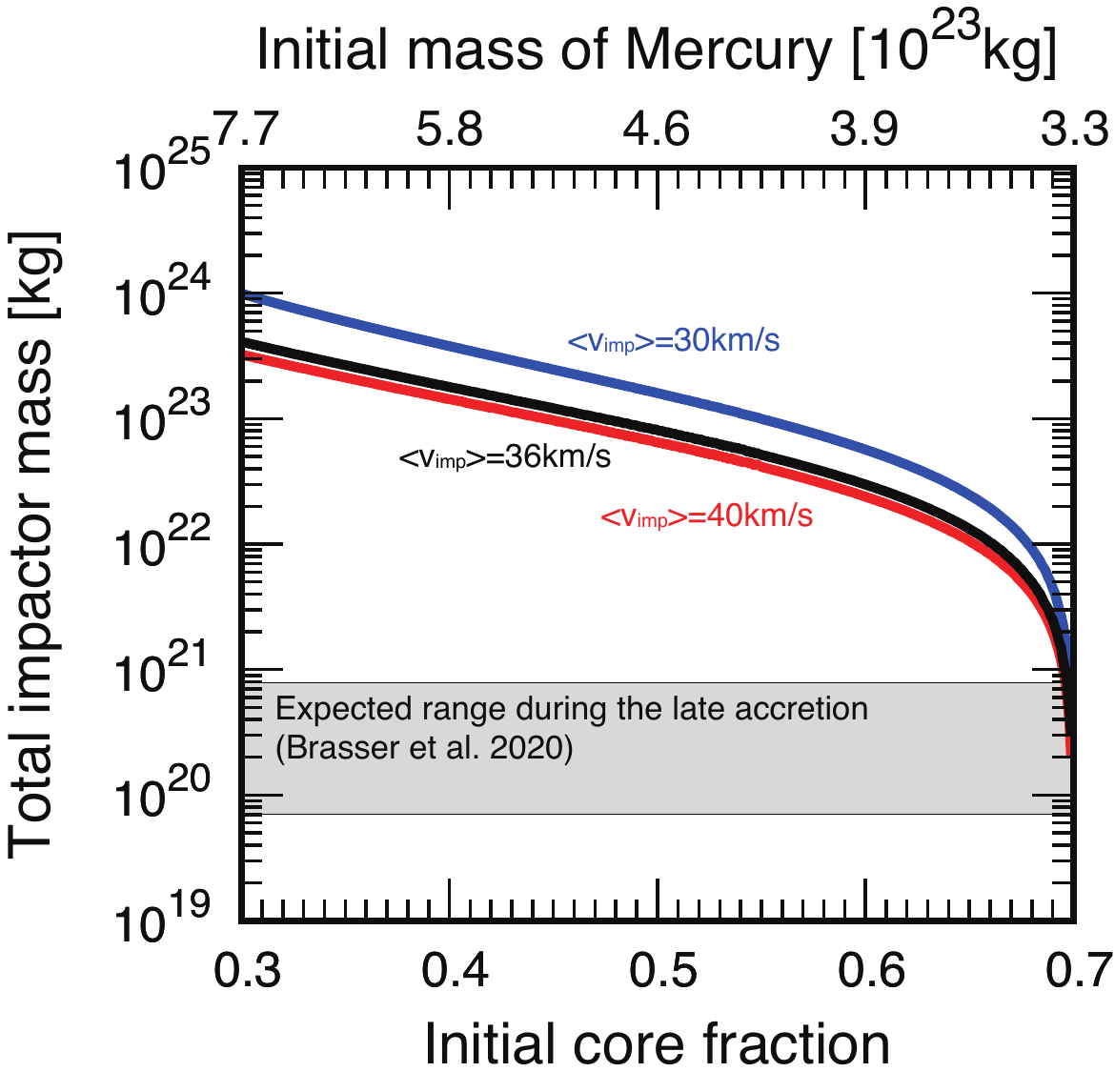}
	\caption{Total mass of the bombardment needed to explain today's core fraction of Mercury ($f_{\rm core}=0.7$) as a function of the initial core fraction assuming the core mass is fixed as that of today. The blue and red solid lines represent cases where impact velocity distributions are normal distributions with $\mu=30$ and 40 km s$^{-1}$ and $\sigma=1$ km s$^{-1}$, respectively. The black solid line represents the case where a modified Rayleigh distribution (Equation \ref{eq_vimp}) is used with $\mu_{\rm imp}=10.31$ km s$^{-1}$ and $\sigma_{\rm imp}=20.81$ km s$^{-1}$ (the mean value is $<v_{\rm imp}>=36$ km s$^{-1}$). The gray shaded region is the expected range by the latest dynamical model \citep{Brasser2020}.}
	\label{Fig_mass_erode_required}
\end{figure*}

Figure \ref{Fig_mass_erode_required} shows the required total bombardment mass, $M_{\rm imp,tot,req}$, to explain Mercury's current core mass fraction beginning from an assumed lower ratio of core:mantle. Assuming proto-Mercury has a chondritic core mass fraction of $f_{\rm core,ini}=0.3$ \citep[e.g.,][]{Ebel2017} with the core mass equal to its current mass with no augmentation to the core from late accretion, the required mass impacting Mercury to account for the present core: mantle value is $M_{\rm imp,tot,req} \sim 3 \times 10^{23} - 10^{24}$ kg for a impact velocity of $v_{\rm imp} = 30-40$ km s$^{-1}$. Considering a giant impact on proto-Mercury of $f_{\rm core}=0.3$, resulting core mass fraction of $f_{\rm core} \sim 0.5$ is relatively a natural outcome \citep[e.g.,][]{Asphaug2014}. In this case, the required total mass of bombardment to increase the core mass fraction from $f_{\rm core,ini}=0.5$ to $f_{\rm core}=0.7$ is $M_{\rm imp,tot,req} \sim 5 \times 10^{22} - 2 \times 10^{23}$ kg for a impact velocity of $v_{\rm imp} = 30-40$ km s$^{-1}$. Interestingly, for both cases of $f_{\rm core,ini} = 0.3$ and $0.5$, the required total mass of the bombardment is much larger than that expected from late accretion in different planet formation scenarios (Table \ref{table_model}; $M_{\rm imp,tot,req} < 10^{21}$ kg; green shaded area in Figure \ref{Fig_mass_erode_required}).\\

Using the averaged arguments (Section \ref{sec_scaling}) and assuming today's size of Mercury, the cumulative depth of the mass escape $D_{\rm esc} = M_{\rm esc,tot}/4\pi \rho R_{\rm Mer}^{2}$ from the surface of Mercury (i.e., the degree of crustal erosion) by late accretion with its total mass of $M_{\rm imp,tot}$ is estimated by using Equation \ref{eq_ero_1} as
%
\begin{eqnarray}
	D_{\rm esc} \simeq \frac{M_{\rm esc,tot}}{4\pi \rho R_{\rm Mer}^{2}} \simeq 1{\rm km} \times \left( \frac{v_{\rm imp}}{36 {\rm \hspace{0.1em} km \hspace{0.15em} s^{-1}}} \right)^{2.2} \left( \frac{M_{\rm imp,tot}}{10^{20} {\rm \hspace{0.1em} kg} }\right)
\label{eq_Dero}
\end{eqnarray}

\noindent where we assume $\rho=3000$ kg m$^{-3}$ for rocky crust and mantle and $R_{\rm Mer}=2400$ km, respectively. $M_{\rm esc,tot}$ is the total escape mass of Mercury's material and we use Equation \ref{eq_ero_1} for $v_{\rm imp} < 12 v_{\rm esc}$, which is valid for $v_{\rm imp} < 52$ km s$^{-1}$ for a value of Mercury's escape velocity of $v_{\rm esc}=4.3$ km s$^{-1}$. Accretion of the impactor material is not considered, but it is negligible compared to the mass escape (Section \ref{sec_erosion}). For $v_{\rm imp} = 30-40$ km s$^{-1}$ and $M_{\rm imp,tot} = 8 \times 10^{18} - 8 \times 10^{20}$ kg (Table \ref{table_model}), the primordial crust of Mercury could be removed by $D_{\rm esc} \sim 50$ m $-10$ km from the surface depending on the impact velocity and the total impactor's mass during late accretion.\\

We conclude that the impact process during late accretion and after Mercury fully formed is generally unable to produce the current high core-to-mantle ratio via impact-induced mass escape of its mantle. Thus, for the rest of this paper, we fix the mass of Mercury as that of today, and we focus on physical outcomes of impact bombardment that would characterize Mercury's surface.\\

\section{What is the fate of impactors striking Mercury?}
\label{sec_question2}
Mercury's surface is heavily cratered from late accretion \citep[see also][]{Mojzsis2018}. To qualitatively and quantitatively understand the fate of impactors on Mercury's surface, we estimate surface expressions of cratering (Section \ref{sec_crater}), the mass of delivery of impactors' material to Mercury as a late veneer (Section \ref{sec_accretion}), and degree of melting and vaporization of the impactors (Section \ref{sec_melting}).\\

In this section, we consider the total impactor's mass as a parameter, and we continue our Monte Carlo simulations until it exceeds a given total mass to obtain statistical values. The total impactor's mass is randomly selected 10,000 times in a logarithmic spaced manner between $M_{\rm imp,tot}=3 \times 10^{19} - 3 \times 10^{21}$ kg which nearly covers the range of interest here (Table \ref{table_model}). Note that we neglect the change in the mass of Mercury and fix the planet's mass to its present value (see the reasoning in Section \ref{sec_erosion}).\\

\subsection{Cratering on Mercury during late accretion} \label{sec_crater}

\begin{table}[width=.95\linewidth,cols=4,pos=t]
\caption{$a$ and $b$ for Equation \ref{eq_Stot} and $\gamma$ for Equation \ref{eq_Macctot}. The dispersions cover $\sim$70\% of $a$ and $\gamma$ obtained from our Monte Carlo simulations.}
\label{table_equation}

\begin{tabular*}{\tblwidth}{@{} LLLL@{} }
\toprule
$ $ & $ <v_{\rm imp}>=30$ km s$^{-1}$ & $<v_{\rm imp}>=36$ km s$^{-1}$ & $<v_{\rm imp}>=40$ km s$^{-1}$ \\
\midrule

a &  $1.79^{+0.66}_{-0.88}$ & $2.15^{+0.76}_{-1.09}$ & $2.45^{+0.84}_{-1.22}$ \\
b & 0.9 & 0.9 & 0.9 \\

\midrule

$\gamma$ & $0.48 \pm 0.12$ & $0.44 \pm 0.12$ & $0.39 \pm 0.12$ \\

\bottomrule
\end{tabular*}
\end{table}

Here, we discuss the crater formation during late accretion. To calculate the size of craters, we use the $\pi$-group scaling law and an empirical relationship linking the transient crater diameter $D_{\rm tr}$ with the final crater diameter $D_{\rm f}$ \citep[e.g.,][]{Melosh1989}. The transient crater diameter is estimated as:
%
\begin{eqnarray}
	D_{\rm tr} = \left( \frac{\pi}{6} \right)^{\frac{1}{3}} C_{\rm D} \left( \frac{4\pi}{3} \right)^{-\frac{\beta}{3}} \left( \frac{\rho_{\rm p}}{\rho_{\rm t}} \right)^{\frac{1}{3}} D_{\rm p}^{1-\beta} g_{\rm Mercury}^{-\beta} \left(  v_{\rm imp}\sin \theta \right)^{2\beta} 
\label{eq_Dt}
\end{eqnarray}
\noindent where $C_{\rm D}=1.6$ and $\beta=0.22$ for nonporous rocks \citep{Schmidt1987}, respectively. $g_{\rm Mercury}=3.7$ m s$^{-2}$ is the current acceleration due to gravity of Mercury. We assume the same densities for target $\rho_{\rm t}$ and impactor $\rho_{\rm p}$. To convert from $D_{\rm tr}$ to $D_{\rm f}$ on Mercury, we used two empirical laws. For small craters, the simple crater assumption is valid \citep{Chapman1986} and 
%
\begin{eqnarray}
	D_{\rm f} = 1.25 D_{\rm tr}.
\label{eq_Df1}
\end{eqnarray}
For large complex crater \citep[e.g.,][]{McKinnon1991},
%
\begin{eqnarray}
	D_{\rm f} = 1.2 D_{\rm SC}^{-0.13} D_{\rm tr}^{1.13}
\label{eq_Df2}
\end{eqnarray}
\noindent where $D_{\rm sc}$ is the crater diameter at which the transition from simple to complex craters is considered \citep{Pike1980} and
%
\begin{eqnarray}
	D_{\rm SC} = 15 {\rm km} \times \frac{g_{\rm Moon}}{g_{\rm Mercury}}
\label{eq_Dsc}
\end{eqnarray}
\noindent where $g_{\rm Moon}=1.62$ m s$^{-2}$ is the acceleration due to gravity of the Moon. We use the larger $D_{\rm f}$ calculated from the above two equations and lunar craters as a baseline. We calculate the cumulative cratered area on Mercury during late accretion as: 
%
\begin{eqnarray}
	S_{\rm tot,cra} = \frac{1}{4} \Sigma \pi D_{\rm f}^{2}
\label{eq_Stot}
\end{eqnarray}

\noindent where $\Sigma$ indicates a sum of the craters.\\

\begin{figure*}
	\centering
	  \includegraphics[width=0.8\textwidth]{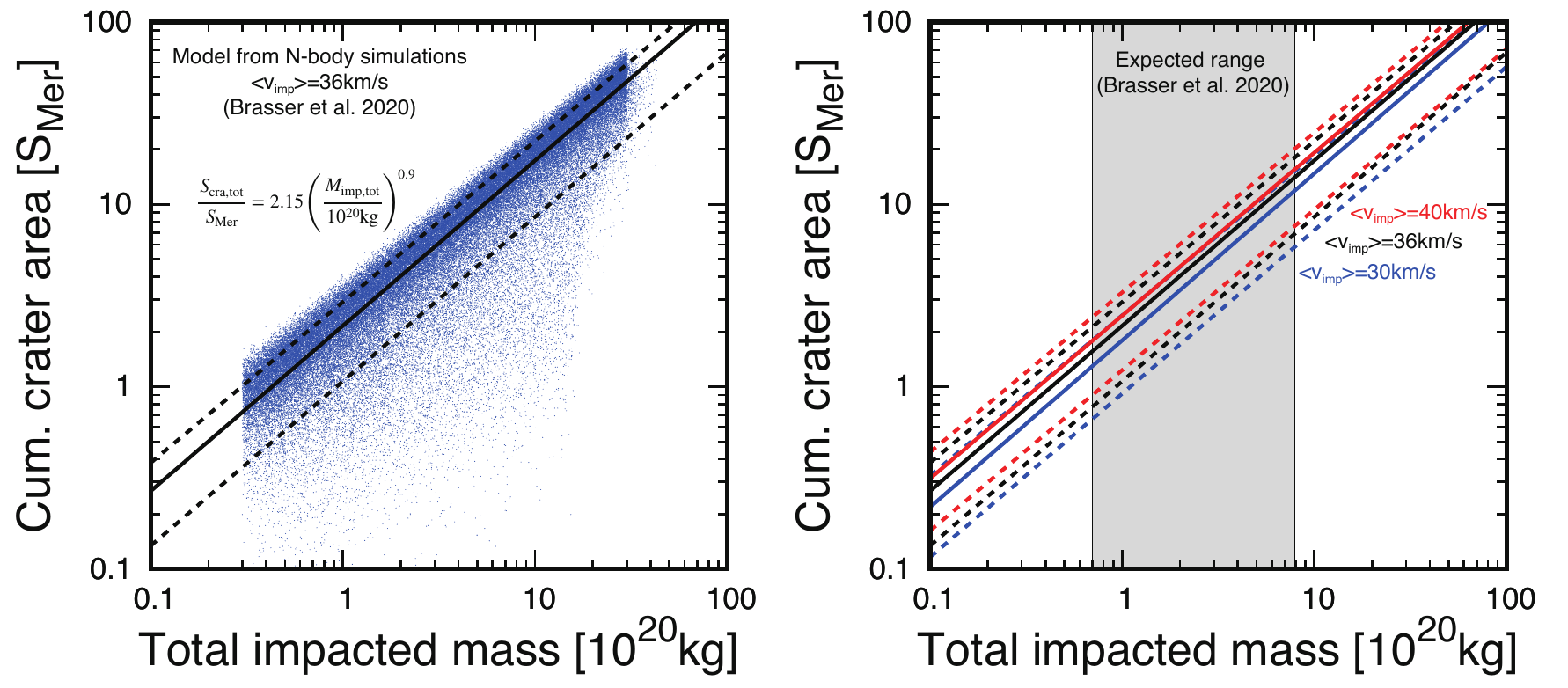}
	\caption{Cumulative crater area as a function of the total impactor mass to Mercury. Left panel: The blue points are results of 100,000 Monte Carlo runs with the total impactor mass between $M_{\rm imp,tot} = 3 \times 10^{19} - 3 \times 10^{21}$ kg for the case of exponential decay model (Section \ref{sec_model}; \cite{Brasser2020}). The solid black line represents the median value fitted using the results between $M_{\rm imp,tot} = 1 \times 10^{20} - 1 \times 10^{21}$ kg. The dashed black lines represent the dispersions in which $\sim$70\% of the accreted mass is covered within the two dashed lines. Right panel: same as the left panel but for different impact velocity distributions. Blue and red lines represent the case where impact velocity distribution is a normal distribution with $\mu=30$ km s$^{-1}$ and $40$ km s$^{-1}$ with $\sigma =1$ km s$^{-1}$, respectively. Black lines are the same as the left panel (the mean value is $<v_{\rm imp}>=36$ km s$^{-1}$). The gray shaded region is the expected range of total impacted mass by the latest dynamical model \citep{Brasser2020}.}
	\label{Fig_crater_area}
\end{figure*}

Figure \ref{Fig_crater_area} shows the degree of resurfacing based on the cumulative crater areas $S_{\rm tot,cra}$ to the surface area of Mercury $S_{\rm Mer}$ as a function of the total mass of the bombardment $M_{\rm imp,tot}$. Using the Bootstrap method, the fitting of our 100,000 Monte Carlo runs (e.g., points in Figure \ref{Fig_crater_area}) provides the $S_{\rm tot,cra}$-$M_{\rm imp,tot}$ scaling relationship for given impact velocity and the total mass of impactors (lines in Figure \ref{Fig_crater_area}) as
%
\begin{eqnarray}
	\frac{S_{\rm tot,cra}}{S_{\rm Mer}} = a \times \left(  \frac{M_{\rm imp,tot}}{10^{20} {\rm kg}} \right)^{b}
\label{eq_Stot}
\end{eqnarray}

\noindent where $a$ and $b$ are the fitting parameters $-$ $a$ depends on the impact velocity, while $b$ is independent on $v_{\rm imp}$ (see Equation \ref{eq_N} and Equations \ref{eq_Dt} - \ref{eq_Dsc}). The resultant $a$ and $b$ for different impact velocity distributions are summarized in Table \ref{table_equation}. Note that, the above arguments implicitly assumed uniform target properties, whereas local properties might play an important role in the cratering process \citep{Marchi2011}.\\

Cratering is an isotropic process. Different models of planet formation suggest Mercury experienced late accretion with $v_{\rm imp} \sim 30-40$ km s$^{-1}$ and $M_{\rm imp,tot} \sim 8 \times 10^{18} - 8 \times 10^{20}$ kg (Table \ref{table_model}). As much as $M_{\rm imp,tot} > 4 \times 10^{19} - 5 \times 10^{19}$ kg bombardment after 4.5 Ga would produce a global resurfacing on Mercury (Figure \ref{Fig_crater_area}; $S_{\rm tot,cra}/S_{\rm Mer} > 1$), which could efficiently erase the older cratering record \citep[see also,][]{Mojzsis2018}. Based on the crater chronology exported from that of the Moon, the surface age of Mercury was estimated \citep[e.g.,][]{Marchi2009,Marchi2011}. The oldest surface age for Mercury was estimated to about $4.0-4.1$ Ga as a result of an extensive global resurfacing \citep{Marchi2013,Werner2014}, which is in accordance with late accretion discussed above. We note that the observed crater size-frequency distribution appears due to the interplay between cratering by impacts and burial of craters by volcanism \citep[see more details in][and references therein]{Marchi2013,Werner2014,Rothery2020}.\\

\subsection{Late veneer on Mercury during impact bombardment}
\label{sec_accretion}
Cratering impact is a process to exchange different materials between impactors and a target \citep[e.g.,][]{Hyodo2020}. In the above section, we studied the mass escape of target material (i.e., target materials that escape from the target gravity by cratering impacts). A fraction of the impactor's material also escapes from the target and the remainder accretes on the target surface during the cratering impacts. In this subsection, we quantitatively investigate the mass of impactors that is implanted on the surface of Mercury during late accretion.\\

\begin{figure*}
	\centering
	  \includegraphics[width=0.8\textwidth]{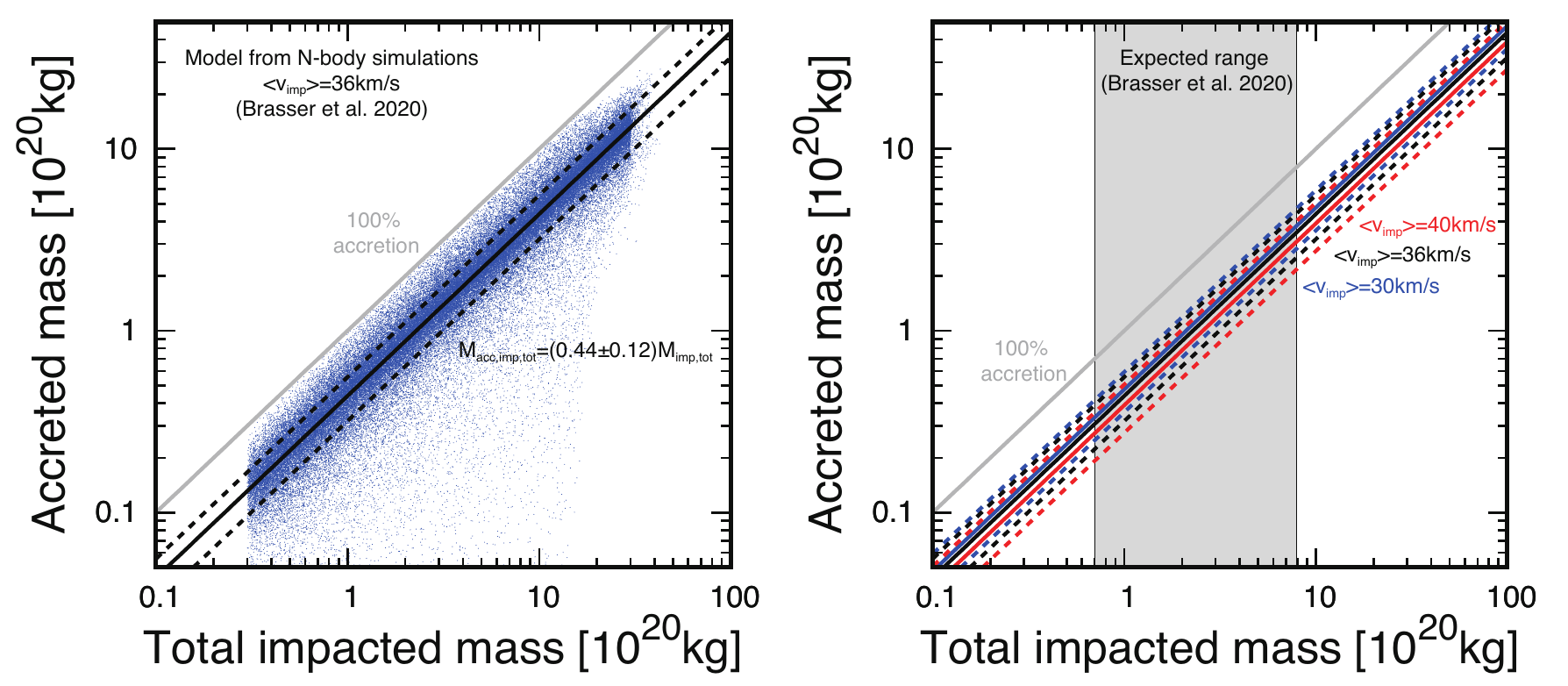}
	\caption{Accretion mass as a function of the total mass of the impactors to Mercury. Left panel: The blue points are results of 100,000 Monte Carlo runs with the total impactor mass between $M_{\rm imp,tot} = 3 \times 10^{19} - 3 \times 10^{21}$ kg for the case of the exponential decay model (Section \ref{sec_model}; \cite{Brasser2020}). The solid black line represents the median value fitted using the results between $M_{\rm imp,tot} = 1 \times 10^{20} - 1 \times 10^{21}$ kg. The dashed black lines represent the dispersions in which $\sim$70\% of the accreted mass is covered within the two dashed lines. The gray line represents the 100\% accretion efficiency. Right panel: the same as the left panel but for the case of the different impact velocity distributions. Blue and red lines represent the case where impact velocity distribution is a normal distribution with $\mu=30$ km s$^{-1}$ and $40$ km s$^{-1}$ with $\sigma=1$ km s$^{-1}$, respectively. Black lines are the same as the left panel (the mean value is $<v_{\rm imp}>=36$ km s$^{-1}$). The gray shaded region is the expected range by the latest dynamical model \citep{Brasser2020}.}
	\label{Fig_mass_accrete}
\end{figure*}

Figure \ref{Fig_mass_accrete} shows the total mass accreted from the impactors on Mercury $M_{\rm acc,imp,tot}$ as a function of the total mass of the impactor. As high as $\sim 40-50$\% of the total mass of the impactors is, on average, incorporated on the surface of Mercury during the cratering impacts (see also Figure \ref{Fig_averaged_masses} middle panel). Using the Bootstrap method, the fitting of our 100,000 Monte Carlo runs derives the $M_{\rm acc,imp,tot} - M_{\rm imp,tot}$ scaling relationship (lines in Figure \ref{Fig_mass_accrete}) as
%
\begin{eqnarray}
	M_{\rm acc,imp,tot} = \gamma M_{\rm imp,tot}
\label{eq_Macctot}
\end{eqnarray}

\noindent where $\gamma$ is the fitting parameter that depends on the impact velocity distribution. The resultant $\gamma$ for different impact velocity distributions is summarized in Table \ref{table_equation}.\\

At each impact, a fraction of the impactor's materials is embedded on the surface of Mercury, and a fraction of the surface materials of the target is ejected. After numerous impacts, the primordial surface materials are mixed with exogenous materials from leftover planetesimals, asteroids, and comets after $4.5$ Ga. As seen in Section \ref{sec_crater}, $M_{\rm imp,tot} > 4 \times 10^{19} - 5 \times 10^{19}$ kg would induce global resurfacing for $v_{\rm imp}=30-40$ km s$^{-1}$. This indicates that a fraction of the previously embedded impactor's materials could be re-ejected by successive impacts. The detailed accretion/re-ejection efficiency depends on each impact because the depth of the penetration of the impactor strongly depends on the impact parameter (see Equation \ref{eq_Dt}). For the conservative estimation, we consider that all the previously embedded impactor's materials are re-ejected every time a global resurfacing takes place ($M_{\rm imp,tot} \sim 4 \times 10^{19} - 5 \times 10^{19}$ kg; Figure \ref{Fig_crater_area}) assuming that the impact parameters do not change during late accretion. This indicates that $M_{\rm acc,imp,tot} \sim 3 \times 10^{18} -1.6 \times 10^{19}$ kg exogenous impactor's materials are at least mixed with the primordial Mercurian surface materials for $M_{\rm imp,tot} = 8 \times 10^{18} - 8 \times 10^{20}$ kg (Table \ref{table_model}).\\

Assuming a global and homogenous deposition of the impactor's materials on the surface of Mercury (no mixing with Mercury's material is assumed), the cumulative thickness of the deposition of the impactor's materials $L_{\rm acc,imp}$ by late accretion $-$ the total mass of the impactors is $M_{\rm imp,tot}$ $-$ is given by using Equation \ref{eq_Macctot} as
%
\begin{eqnarray}
	L_{\rm acc,imp} \simeq \frac{M_{\rm acc,imp,tot}}{4\pi \rho R_{\rm Mer}^{2}} \simeq 460 {\rm m} \times \gamma  \left( \frac{M_{\rm imp,tot}}{10^{20} {\rm kg}} \right)
\label{eq_Limp}
\end{eqnarray}

\noindent where $\rho=3000$ kg m$^{-3}$ is the density of the impactor. For a reference value of $<v_{\rm imp}>=36$ km s$^{-1}$ ($\gamma=0.44$; Table \ref{table_equation}) and for $M_{\rm imp,tot} = 8 \times 10^{18} - 8 \times 10^{20}$ kg (Table \ref{table_model}), $L_{\rm acc,imp} \sim 16$ m $-1.6$ km.\\

In reality, some impacts are more efficient/inefficient at embedding a large/small amount of the impactor's materials deep/shallow into the surface of Mercury by mixing the primordial materials, and it all depends on impact conditions. After the numerous impacts (late accretion), the cumulative outcomes of impacts are averaged out. The composition of the embedded exogenous materials depends on the source of the impactor (planetesimals, asteroids or comets). Detailed dynamical studies that statistically resolve the composition of the impactor during late accretion are necessary. However, more fundamentally, our understanding of planetesimal formation is limited at this moment, and it is challenging to understand the initial compositional distribution of the building blocks of the planets and small bodies.\\

\subsection{Melting and vaporization of impactors during late accretion}
\label{sec_melting}
The accretional environment is unique at Mercury's distance to the Sun (see also Section \ref{sec_bui}). Those impactors originating from a distant location are expected to have high-speed impact velocity to Mercury, which is about $v_{\rm imp} \sim 30-40$ km s$^{-1}$ \citep[e.g.,][]{LeDeuvre2008, Mojzsis2018, Hyodo2018, Brasser2020}. Impacts of small bodies on planetary bodies are a local process of planet's surfaces that increases pressure and entropy. \cite{Mojzsis2018} investigated the thermal effects of Mercury's crust and mantle during late accretion and they reported that significant crustal melting inevitably occurred.\\

Here, we focus on the thermodynamical fate of the materials of impactor accreted by Mercury during late accretion, which is not studied in the previous works. This is a critical first step to understand how the impactor's materials evolve on Mercury's surface and to understand the possible outcome to the surface composition of Mercury. Here, we estimate the degree of impact melting and vaporization for impactors to Mercury. We consider the entropy increase $\Delta S$ during the shock compression followed by impact along the Hugoniot curve and calculate the final state of the entropy, $S_{\rm fin} = S_{\rm ini} + \Delta S$ \citep[see details in][]{Sugita2012,Dauphas2015}. Here, impactors are assumed to be homogeneously shocked, forming the isobaric core, because the decay of the shock pressure is limited for such a small distance (i.e., the impactor size) from the impact point \citep[][]{Pierazzo2000}.\\

Initial entropy is given by $S_{\rm ini}$ that corresponds to a temperature of 293 K for basaltic material \citep{Ahrens1972}. This condition is within the range of the surface temperature of Mercury ($\sim 90-700$ K, which depends on daytime or nighttime). Starting from higher initial temperature results in larger amount of melt/vapor and vice versa \citep{Dauphas2015}. The other parameters used in our calculation are the same as those in \cite{Dauphas2015} (see their appendix). Note that, SPH simulations and the above arguments using the Hugoniot curve do not exactly share the same criteria for the evaporation, whereas both of them consider the same effects from the first principals.\\

\begin{table}[width=0.8\linewidth,cols=4,pos=t]
\caption{The incipient and complete entropies for melting and vaporization \citep{Ahrens1972}.}
\label{table_lever}

\begin{tabular*}{\tblwidth}{@{} LLLL@{} }
\toprule

$S_{\rm im}$ [kJ/K/Kg] &  $S_{\rm cm}$ [kJ/K/Kg] & $S_{\rm iv}$ [kJ/K/Kg] & $S_{\rm cv}$ [kJ/K/Kg] \\
\midrule

2.467 & 2.682 & 3.461 & 7.654 \\

\bottomrule
\end{tabular*}
\end{table}

Using the lever rule, we estimate the melt ($\psi_{\rm mel}$) and vapor ($\psi_{\rm vap}$) fractions, as follows:
%
\begin{eqnarray}
	&\psi_{\rm mel} = 0 \hspace{2em} {\rm for} \hspace{2em} S_{\rm fin} < S_{\rm im}\\
	&\psi_{\rm mel} = \frac{S_{\rm fin} - S_{\rm im}}{S_{\rm cm} - S_{\rm im}} \hspace{2em} {\rm for} \hspace{2em} S_{\rm im} < S_{\rm fin} < S_{\rm cm}\\
	&\psi_{\rm mel} = 1 \hspace{2em} {\rm for} \hspace{2em} S_{\rm fin} > S_{\rm cm}
\label{eq_psi_mel}
\end{eqnarray}
%
\begin{eqnarray}
	&\psi_{\rm vap} = 0 \hspace{2em} {\rm for} \hspace{2em} S_{\rm fin} < S_{\rm iv}\\
	&\psi_{\rm vap} = \frac{S_{\rm fin} - S_{\rm iv}}{S_{\rm cv} - S_{\rm iv}} \hspace{2em} {\rm for} \hspace{2em} S_{\rm iv} < S_{\rm fin} < S_{\rm cv}\\
	&\psi_{\rm vap} = 1 \hspace{2em} {\rm for} \hspace{2em} S_{\rm fin} > S_{\rm cv}
\label{eq_psi_vap}
\end{eqnarray}
\noindent where $S_{\rm im}$, $S_{\rm cm}$ and $S_{\rm iv}$, $S_{\rm cv}$ are the incipient and complete entropies for melting and vaporization for basaltic materials, respectively (Table \ref{table_lever}). When $0 < \psi_{\rm vap} < 1$, the (effective) melt fraction is corrected as $\psi_{\rm mel,eff} = \psi_{\rm mel} - \psi_{\rm vap}$.\\

\begin{figure*}
	\centering
	  \includegraphics[width=0.5\textwidth]{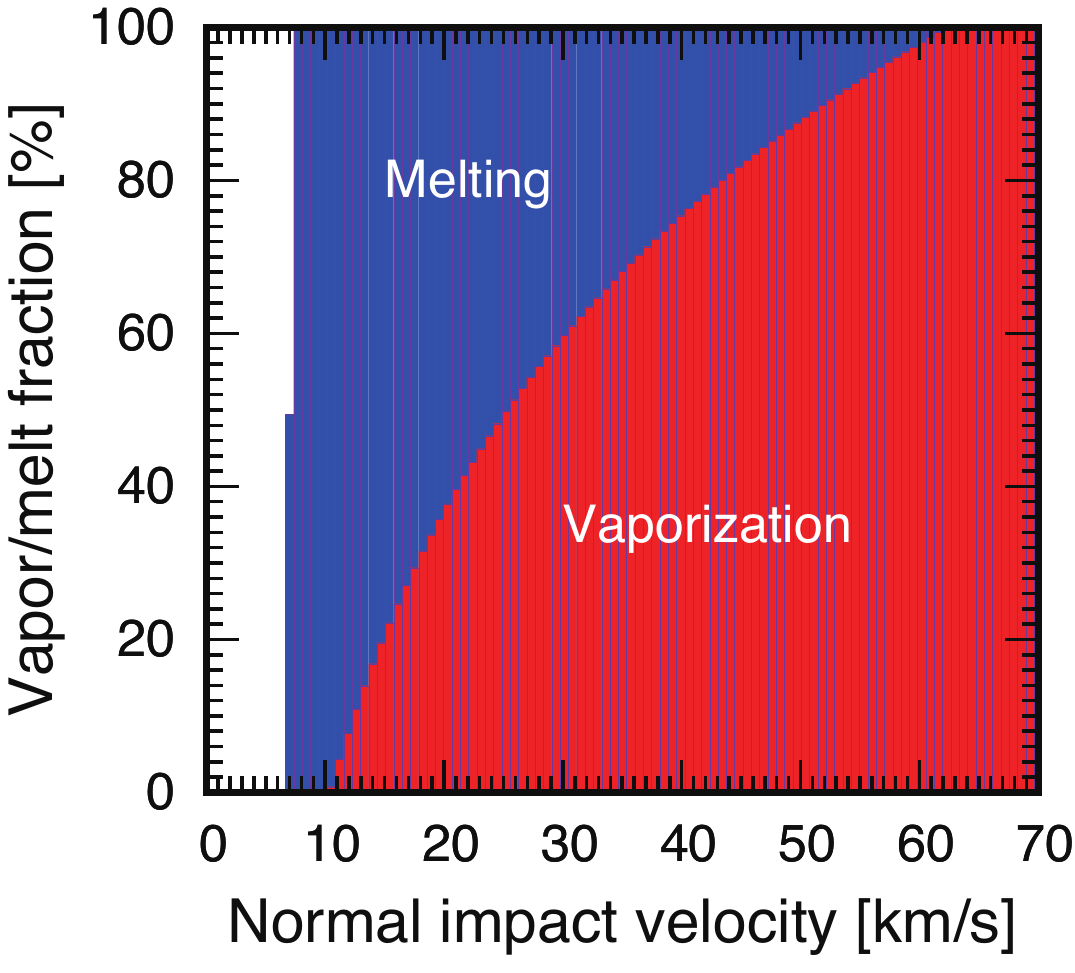}
	\caption{Fraction of impact-vaporization/melting of the impactor as a function of the normal component of the impact velocity. Red field indicates the fraction of the vaporization. Blue field indicates the fraction of the melting. The most probable impact angle is $\theta=45$ degrees \citep{Shoemaker1962}. For the cases of $<v_{\rm imp}>=30$ and $40$ km s$^{-1}$, the normal component of the impact velocities are $v_{\rm imp,nor}=<v_{\rm imp}> \sin(\theta)=21$ and $28$ km s$^{-1}$, respectively. For the case of the statistical value from the latest $N$-body simulations, $<v_{\rm imp}>=36$ km s$^{-1}$ \citep{Brasser2020} and $v_{\rm imp,nor}=25$ km s$^{-1}$. For $v_{\rm imp,nor}<11$ km s$^{-1}$, only impact melting takes place. For $v_{\rm imp,nor}<7$ km s$^{-1}$, there is no melting and vaporization. A sharp change in the melting fraction at $v_{\rm imp,nor} \sim 7$ km s$^{-1}$ is due to the small difference between $S_{\rm im}$ and $S_{\rm cm}$ (Table \ref{table_lever}).}
	\label{Fig_frac_vap_mel}
\end{figure*}

Figure \ref{Fig_frac_vap_mel} shows the melt and vapor fractions for a given impact velocity. Here, the normal component of the impact velocity $v_{\rm imp,nor}$ is considered because it mainly controls the melting and vaporization \citep{Pierazzo2000}. For the typical impact velocity ranging between $v_{\rm imp} \sim 30-40$ km s$^{-1}$ \citep{Mojzsis2018, LeDeuvre2008, Brasser2020} with the most probable impact angle of $\theta=45$ degrees \citep{Shoemaker1962}, $v_{\rm imp,nor} \sim 21-28$ km s$^{-1}$. Figure \ref{Fig_frac_vap_mel} indicates that $\sim 40-50$\% of the mass fraction of the impactor is vaporized and $\sim 50-60$\% of the impactor's materials are melted. These melted and vaporized impactor's materials are embedded on the surface of Mercury as a late veneer (see also Section \ref{sec_accretion}). \\

More detailed and complex studies are required to understand the long-term evolution of the melted/vaporized impactor's materials. Late accretion is an ensemble of small impactors $-$ they could have a variety of compositions and impact parameters depending on their source(s) $-$ and thus it produces a distinctive local change in chemical and thermal aspects on the surface of Mercury. The local magma ponds and vapor produced by small impacts considered here would cool down much quickly compared to the case of the global magma ocean state produced by a giant impacts \citep[e.g.,][]{Benz2007}. As the impactor's materials are melted or vaporized, the comparable amount of the target materials, at least, is melted and vaporized \citep[see more details in][]{Mojzsis2018}. These different materials might mix homogeneously and form new mixtures, or the external materials from impactors, including volatile-bearing materials, could be heterogeneously implanted on the surface of Mercury. Either way, late accretion is a process to deliver exogenous materials to Mercury's surface and could dictate to some degree the composition of the surface of Mercury. This is one of the critical aspects to understand the observed surface morphological, chemical, and compositional characteristics of Mercury.\\

\section{Erosion of the building blocks of Mercury by cratering impacts}
\label{sec_bui}
In section \ref{sec_erosion}, we demonstrated that late accretion hardly produces a significant amount of mass escape of the mantle material because a fully formed Mercury is already too massive for the impact ejecta to efficiently escape from its gravity ($v_{\rm esc}>4.3$ km s$^{-1}$).\\

\begin{figure*}
	\centering
	  \includegraphics[width=0.5\textwidth]{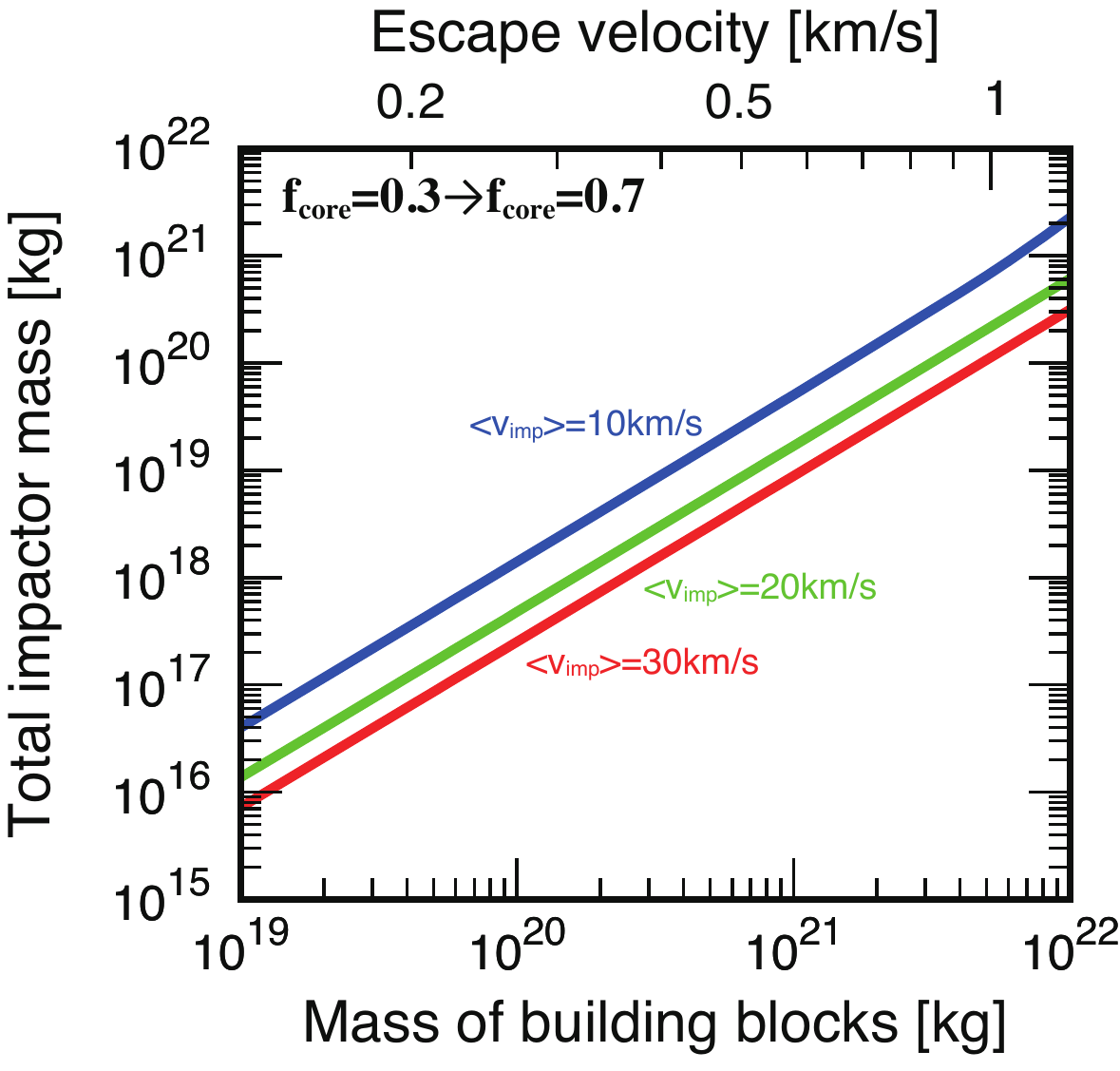}
	\caption{Total bombardment mass required to remove mantle of the putative building blocks of planets from $f_{\rm core}=0.3$ to $f_{\rm core}=0.7$. The blue, green and red solid lines represent cases where impact velocity distributions are the normal distributions with $\mu=$10, 20, 30 km s$^{-1}$ and $\sigma=1$ km s$^{-1}$, respectively. As comparisons, the mass of Vesta is $\sim 3\times10^{20}$ kg and that of Ceres is $\sim 10^{21}$ kg, respectively. Mass of Mercury is $\simeq 3.3 \times 10^{23}$ kg and its escape velocity is $\simeq 4.3$ km s$^{-1}$. }
	\label{Fig_mass_erode_building_blocks}
\end{figure*}

Here, alternatively, we investigate the case where small differentiated building blocks of planets experience impact bombardment (bottom panel of Figure \ref{Fig_summary}). Figure \ref{Fig_mass_erode_building_blocks} shows the required mass, $M_{\rm imp,tot}$, to erode the silicate portion of differentiated bodies so that the core mass fraction increases from $f_{\rm core}=0.3$ to $f_{\rm core}=0.7$ as a function of the mass of the building blocks $M_{\rm bui}$. As the mass of the building blocks decreases, the ratio of $M_{\rm imp,tot}/M_{\rm bui}$ steeply decreases from $M_{\rm imp,tot}/M_{\rm bui} \sim 10^{-1}$ to $M_{\rm imp,tot}/M_{\rm bui} \sim 10^{-3}$ for $<v_{\rm imp}>=10$ km s$^{-1}$ as $M_{\rm bui}=10^{22}$ kg to $M_{\rm bui}=10^{19}$ kg. This is because the escape mass is a strong function of the escape velocity of the target (Equations \ref{eq_ero_1} and \ref{eq_ero_2}; see also \cite{Hyodo2020}).\\

The above results indicate that if the formation of the building blocks were fast and efficiently differentiated driven by heat from the radioactive decay of the short-lived radioisotopes (in particular $^{26}$Al) in the early solar system \citep[e.g.,][]{Grimm1993}, the mantles of the building blocks of Mercury could be significantly removed by cratering impacts, resulting in a large core fraction from which Mercury could form through successive giant impacts and assimilation of these "naked" cores. The Keplerian velocity is $v_{\rm K} \sim 50$ km s$^{-1}$ at Mercury's orbital distance. The impact velocity is given by $\sqrt{v_{\rm esc}^2+v_{\rm ran}^2}$, with $v_{\rm ran} \sim \sqrt{e^2 + i^2}v_{\rm K}$, where $e$ and $i$ are the eccentricity and inclination, respectively (assumed to both be fairly low). The escape velocity of the building blocks is low ($v_{\rm esc} < 1$ km s$^{-1}$ for $M_{\rm bui}<10^{22}$ kg). In contrast, much less massive impactors may have high eccentricities (a high random velocity) excited gravitationally by distant large objects such as proto-Earth and/or proto-Venus. An impact velocity of $<v_{\rm imp}> \sim 10$ km s$^{-1}$ occurs when $e \sim 0.2$ and nearby or distant lager embryos can excite the orbits of small bodies. The situation is different at the location of the Earth. The orbital velocity of the Earth is $\sim 30$ km $^{-1}$, and $e > 0.33$ is required to have $<v_{\rm imp}>=10$ km s$^{-1}$ on small building blocks. In the classical in-situ accretion scenario \citep[e.g.,][]{Safronov1972, Hayashi1985}, this could be difficult to achieve given that the typical maximum excitation of the eccentricity is $e_{\rm max} \sim v_{\rm esc}/v_{\rm K}$ and the isolation mass of planetary embryos is an increasing function of heliocentric distance. An instability of the early solar system \citep[e.g.,][]{Mojzsis2019,Clement2019b} or a runaway accretion of planetesimals in a narrow annulus \citep[e.g.,][]{Hansen2009, Hyodo2019a} might have an excited accretional environment.\\

The effects of collisional erosion was also emphasized in embryo-embryo or embryo-planetesimal collisions \citep[$D>100$ km;][]{Carter2018} whose size is much larger than those considered in this study. They found that the surface materials could be preferentially stripped from embryos as they accrete through the collisions between the similar sized bodies during planet formation (mass ratio $> 0.01$). Their results also indicate that the surface materials could be selectively reduced compared to the core materials resulting in a large core fraction. Thus, different scales of planetary impacts $-$ cratering impact, embryo-embryo collision, and giant impact $-$ might have played a role to erode Mercury's mantle.\\

The ejected materials could be efficiently removed by Poynting-Robertson drag \citep[e.g.,][]{Burns1979, Benz2007, Gladman2009}, radiation pressure \citep[e.g.,][]{Burns1979,Hyodo2018b} and/or strong solar wind at the time of the early solar system \citep[e.g.,][]{Spalding2020}. Detailed studies are required to constrain the validity of the above scenarios further. This requires studies of the planetesimal formation, differentiation, and the accretional process in the early stage of the planet formation.\\

\section{Conclusions}
\label{sec_conclusion}
Impacts are a fundamental process by which planets grow and are modified. Stochastic giant impacts on the terrestrial bodies mechanically and thermally affect a large portion of the planet's surface \citep[e.g.,][]{Benz1988, Nakajima2015, Hyodo2018b}. Contrarily, small impacts, namely the cratering impacts, affect only a small area of the planet's surface and are much more frequent \citep[e.g.,][]{Melosh1989impact,Melosh2011}.\\

In this work, using analytical and Monte-Carlo approaches combined with the scaling laws for the escape mass of the target material and the accretion mass of the impactor material during the cratering impacts \citep{Hyodo2020}, we studied (1) whether late accretion significantly erodes Mercury (Section \ref{sec_question1}), and (2) the fate of the impactors to Mercury during late accretion (Section \ref{sec_question2}). Considering the uncertainties in late accretion to Mercury, we developed scaling laws for the following parameters as a function of impact velocity and total mass of late accretion: (1) depth of crustal erosion (Equation \ref{eq_Dero}), (2) the degree of resurfacing (Equation \ref{eq_Stot}), and (3) mass accreted from impactor material (Equation \ref{eq_Macctot}).\\

Existing dynamical models of planet formation indicated that late accretion ($M_{\rm imp,tot} \sim 8 \times 10^{18} - 8 \times 10^{20}$ kg; Table \ref{table_model}) took place on Mercury with a typical impact velocity of $v_{\rm imp} \sim 30-40$ km s$^{-1}$ after 4.5 Ga \citep[e.g.,][]{LeDeuvre2008, Marchi2013, Mojzsis2018, Mojzsis2019, Brasser2020}. For this parameter range, analytical arguments (Section \ref{sec_erosion}) showed that late accretion could remove Mercury's surface crust by $D_{\rm esc} \sim 50$ m $-10$ km, but the change in its core fraction was negligible. More than $10^{23}$ kg of bombardment is required to remove enough of the mantle to produce the current core mass fraction of $f_{\rm core} \sim 0.7$ from an initially chondritic value of $f_{\rm core} \sim 0.3$. However, the dynamical model indicates $<10^{21}$ kg bombardment occurred (Figure \ref{Fig_mass_erode_required}).\\

Alternatively, our results indicated that the silicate mantles of the assumed differentiated building blocks of Mercury could be effectively eroded, if they formed quickly and were efficiently differentiated in the early solar system, assuming they are $10^{19} - 10^{21}$ kg and their escape velocities $v_{\rm esc} \sim 0.1-0.5$ km s$^{-1}$ (Section \ref{sec_bui}; bottom panel of Figure \ref{Fig_summary}). For example, a total impacted mass of $\sim 10^{16}$ kg could change $10^{19}$ kg building blocks (radius of $\sim 100$ km) from $f_{\rm core} = 0.3$ to $f_{\rm core} = 0.7$ with $v_{\rm imp} \sim 10$ km s$^{-1}$ (Figure \ref{Fig_mass_erode_building_blocks}). Detailed studies are required to constrain the validity of this scenario further.\\

We showed that an intensive cratering during late accretion could globally resurface the oldest geological record of Mercury (Section \ref{sec_crater}; top panel of Figure \ref{Fig_summary} and see also \cite{Mojzsis2018}). Our Monte Carlo simulations indicated that $M_{\rm imp,tot} > 4 \times 10^{19} - 5 \times 10^{19}$ kg with $v_{\rm imp} \sim 30-40$ km s$^{-1}$ on average globally resurfaced Mercury's surface (Figure \ref{Fig_crater_area}). At the same time the cratering takes place, a fraction of the impactor's materials is embedded on the surface of Mercury (Section \ref{sec_accretion}). Because previously accreted impactor's materials could be re-ejected by successive cratering impacts, we estimated that $M_{\rm acc,imp,tot} \sim 3 \times 10^{18} - 1.6 \times 10^{19}$ kg exogenous impactor's materials were at least mixed with the primordial Mercurian surface materials as a late veneer (Figure \ref{Fig_mass_accrete}).\\

The impactor's materials were completely melted or vaporized $-$ about $40-50$\% of the mass fraction of the impactors was vaporized and the rest was melted (Figure \ref{Fig_frac_vap_mel} in Section \ref{sec_melting}). As well as impactors, a comparable amount of Mercury's surface is, at least, melted and vaporized \citep[see][]{Mojzsis2018}. Further detailed studies are required to understand the final state of the phase-changed material. The impactor, including the volatile-bearing materials, could equilibrate with the primordial Mercurian materials or be heterogeneously buried on the surface of Mercury as exogenic materials. This process could characterize the surface morphology and composition of Mercury (top panel of Figure \ref{Fig_summary}).\\

In conclusion, late accretion seems an inevitable dynamical process at the very last stage of the planet formation, and it affects Mercury's surface in both mechanical and thermal aspects. Late accretion is a process to deliver exogenic impactor's materials from a distant location to Mercury. The coverage for Mercury from MESSENGER was largely limited to the northern hemisphere although there were important findings. Detailed measurements of small craters on Mercury would shed more light on the crater chronology. The global mapping of the surface mineralogy and composition, including a better characterization of the volatile species, would improve the interpretation of late accretion and the resultant imprints of impactor's materials on the surface of Mercury. The Mercury Planetary Orbiter (MPO) onboard the ESA-led BepiColombo mission, such as MERTIS (MErcury Radiometer and Thermal Infrared Spectrometer), SIMBIO-SYS (Spectrometer and Imagers for MPO BepiColombo-Integrated Observatory SYStem),  MIXS (Mercury Imaging X-ray Spectrometer), MGNS (Mercury Gamma-ray and Neutron Spectrometer), and BELA (BEpiColombo Laser Altimeter), will provide the first comprehensive view of Mercury's entire surface. We expect that the further interplay between our theoretical results and the detailed surface observations of Mercury, including the BepiColombo mission, will lead us to a better understanding of Mercury's origin and evolution.\\

We thank anonymous referees for their constructive comments that helped greatly improve the manuscript. R.H. acknowledges the financial support of JSPS Grants-in-Aid (JP17J01269, 18K13600) and JAXA's ITYF supports. RB acknowledges financial assistance from JSPS Shingakujutsu Kobo (JP19H05071). H.G. is supported by JSPS Kakenhi Grant (JP17H02990) and MEXT Kakenhi Grant (JP17H06457).

\bibliography{Mercury}


\end{document}